\documentclass[twocolumn,preprintnumbers,showpacs,pre,floatfix,superscriptaddress,amsmath,amssymb]{revtex4}
\usepackage{epsfig}
\usepackage{subfigure}
\usepackage{amsmath}
\usepackage{color}
\usepackage{amssymb}
\usepackage{setspace}
\usepackage{graphicx}
\usepackage{dcolumn}
\usepackage{bm}
\usepackage{times}
\usepackage{enumerate}
\input epsf

\begin{document}

\author{Per Sebastian Skardal}
\email{skardal@colorado.edu} 
\affiliation{Department of Applied Mathematics, University of Colorado at Boulder, Colorado 80309, USA}

\author{Edward Ott}
\affiliation{Institute for Research in Electronics and Applied Physics, University of Maryland, College Park, Maryland 20742, USA}

\author{Juan G. Restrepo} 
\affiliation{Department of Applied Mathematics, University of Colorado at Boulder, Colorado 80309, USA}

\title{Cluster Synchrony in Systems of Coupled Phase Oscillators with Higher-Order Coupling}


\begin{abstract}
We study the phenomenon of cluster synchrony that occurs in ensembles of coupled phase oscillators when higher-order modes dominate the coupling between oscillators. For the first time, we develop a complete analytic description of the dynamics in the limit of a large number of oscillators and use it to quantify the degree of cluster synchrony, cluster asymmetry, and switching. We use a variation of the recent dimensionality-reduction technique of Ott and Antonsen (Chaos {\bf 18}, 037113 (2008)) and find an analytic description of the degree of cluster synchrony valid on a globally attracting manifold. Shaped by this manifold, there is an infinite family of steady-state distributions of oscillators, resulting in a high degree of multi-stability in the cluster asymmetry. We also show how through external forcing the degree of asymmetry can be controlled, and suggest that systems displaying cluster synchrony can be used to encode and store data. 
\end{abstract}

\pacs{05.45.Xt, 05.90.+m}

\maketitle

\section{Introduction}

Large systems of coupled oscillators occur in many examples throughout science and nature and serve as a basic model for emergent collective behavior. Examples include synchronized flashing of fireflies \cite{FireFly}, cardiac pacemaker cells \cite{Pacemaker}, walker-induced oscillations of the Millennium Bridge \cite{Millenium}, and circadian rhythms in mammals \cite{Circadian}. Under certain conditions, these limit cycle oscillators can be approximately described entirely in terms of their phase angles $\theta$. Kuramoto showed \cite{Kuramoto1} that the evolution of the phases in an ensemble of $N$ weakly coupled oscillators obeys
\begin{equation}\label{eqModelGeneral}
\dot{\theta}_n=\omega_n+\sum_{m=1}^N H_{nm}(\theta_m-\theta_n),
\end{equation}
where $\theta_n$ and $\omega_n$ are the phase and intrinsic frequency of oscillator $n$, and $H_{nm}$ is a $2\pi$-periodic function. The choice of $H_{nm}(\theta)=(K/N)\sin(\theta)$, which leads to the Kuramoto model \cite{Kuramoto1}, is well motivated because the expansion of several coupled oscillators about a Hopf bifurcation generically leads to sinusoidal coupling. This choice has also become a paradigm for the study of emergence of synchrony in coupled heterogeneous oscillators \cite{Strogatz1}. Many generalizations of the Kuramoto model have been studied. Some examples are systems that account for time-delay \cite{Lee1}, network structure \cite{Pikovsky1}, non-local coupling \cite{Martens1}, external forcing \cite{Childs1}, non-sinusoidal coupling \cite{Daido1}, Josephson junction circuits \cite{Marvel1}, coupled excitable oscillators \cite{Alonso1}, bimodal distributions of oscillator frequencies \cite{Martens2}, phase resetting \cite{Levajic1}, time-dependent connectivity \cite{So1}, noise \cite{Nagai1}, and communities of coupled oscillators \cite{Kawamura1}. Recent analytical work \cite{OA1,OA2,Ott1} (in particular the Ott-Antonsen (OA) ansatz \cite{OA1}) has allowed for the simplification of the analysis of these systems to the study of reduced low-dimensional equations and made many of these systems analytically tractable. 

\begin{figure*}[t]
\centering
\addtolength{\belowcaptionskip}{-6mm}
\subfigure{
\epsfig{file =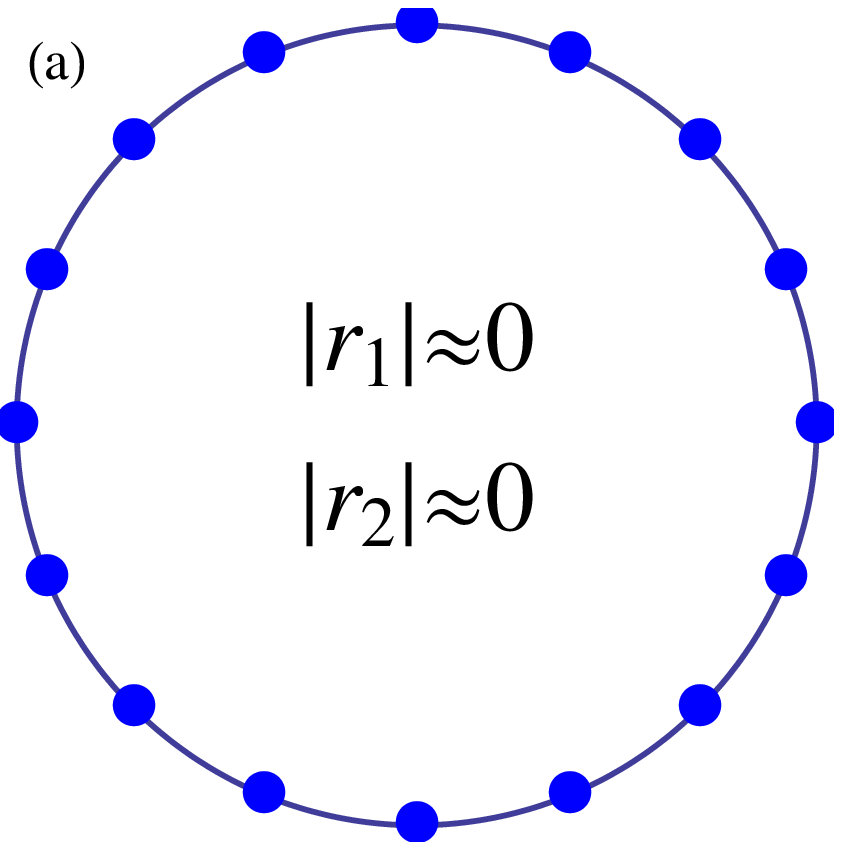, clip =,width=0.28\linewidth } \hskip0.05\linewidth \epsfig{file =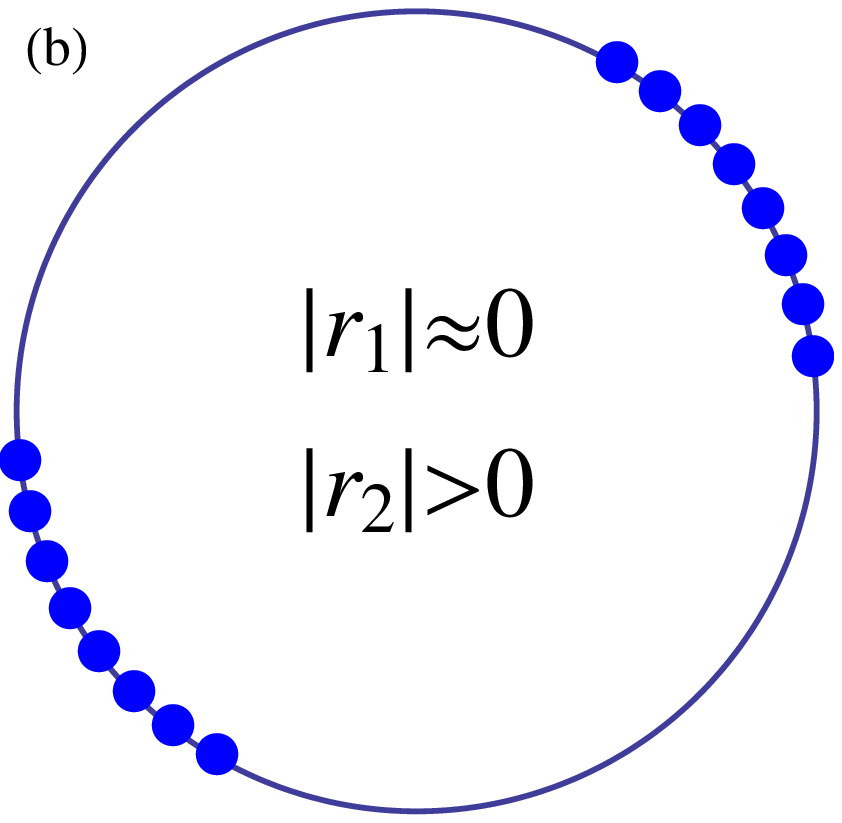, clip =,width=0.28\linewidth } \hskip0.05\linewidth  \epsfig{file =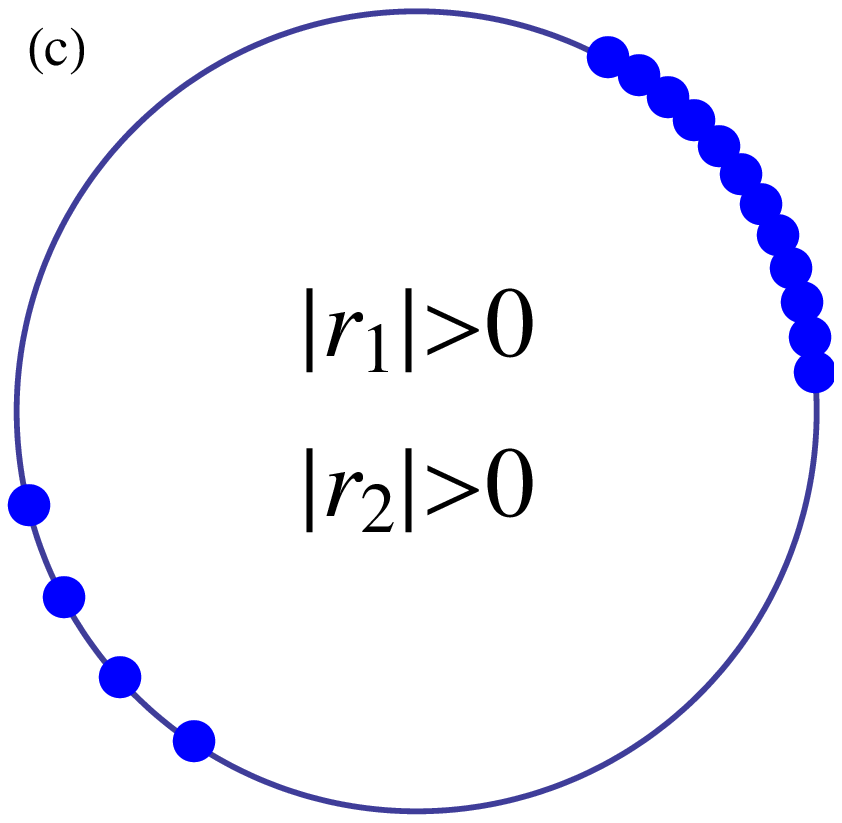, clip =,width=0.28\linewidth }
}
\subfigure{
\epsfig{file =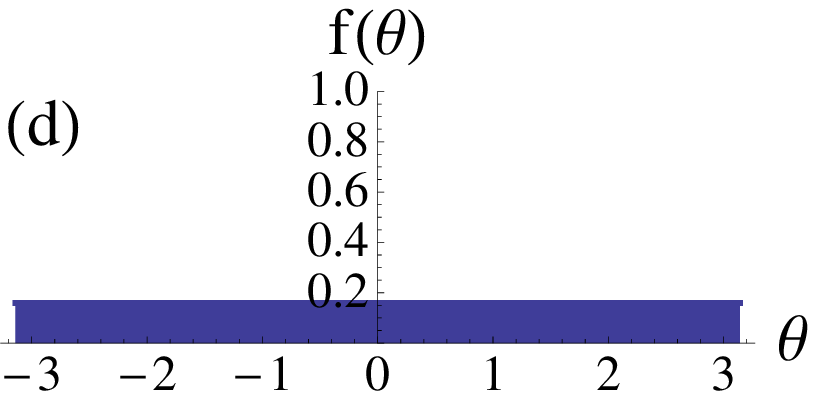, clip =,width=0.28\linewidth } \hskip0.05\linewidth \epsfig{file =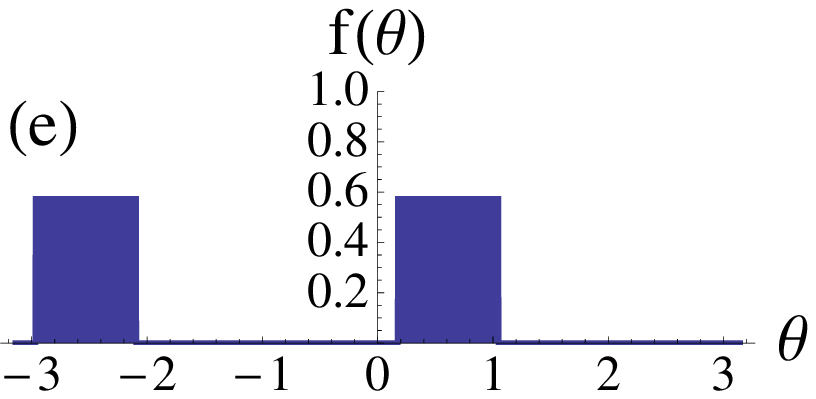, clip =,width=0.28\linewidth } \hskip0.05\linewidth \epsfig{file =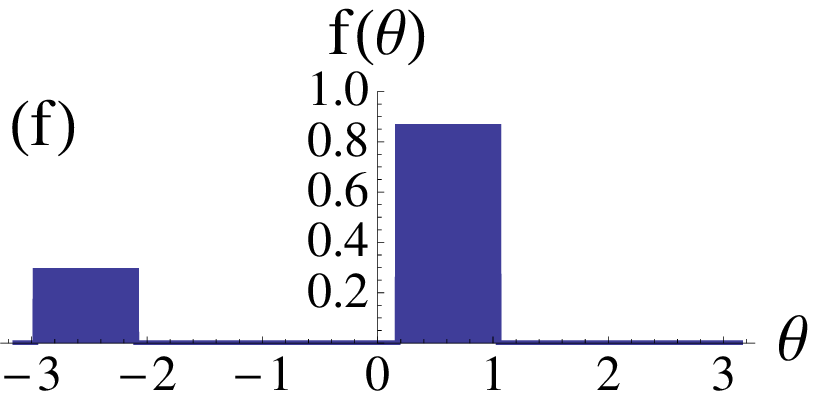, clip =,width=0.28\linewidth }
}
\caption{(Color online) Example oscillator configurations of a system with (a) no synchrony ($|r_1|,|r_2|\approx0$), (b) symmetric ($|r_1|\approx0$) cluster synchrony ($|r_2|>0$), and (c) asymmetric ($|r_1|>0$) cluster synchrony ($|r_2|>0$) and corresponding density functions (d), (e), and (f).} \label{cartoon}
\end{figure*}

While the choice $H_{nm}(\theta)=(K/N)\sin(\theta)$ that yields the Kuramoto model is the simplest, describes many situations of interest, and has the advantage of being analytically tractable, more general choices can result in additional dynamical features. If there are higher harmonics in $H_{nm}$, but the sinusoidal term is dominant, there is a transition to synchrony as in the Kuramoto model as the coupling between the oscillators increases \cite{Daido1}. In this case, the synchronous state is characterized by the phases of a large fraction of the oscillators clustering around a common phase. When higher harmonic terms are dominant, however, the synchronous state is characterized by the formation of multiple synchronized groups (or ``clusters'') of oscillators, each with a common phase \cite{Hansel1}. This phenomenon has also been called multibranch entrainment in previous work \cite{Daido2}. Cluster synchrony occurs in many applications in nature, including networks of neuronal, photochemical, and electrochemical oscillators \cite{Ermentrout1,Taylor1,Kiss1}, as well as genetic networks \cite{Zhang1}. In this paper we will study Eq.~(\ref{eqModelGeneral}) with 
\begin{equation}\label{eqH}
H_{nm}(\theta)=\frac{K}{N}\sin(q\theta),
\end{equation}
 for integer $q\ge2$, which, as we will see, leads generically to the formation of $q$ clusters. There are various experimental and theoretical motivations for the study of this model. In Ref. \cite{Taylor1}, experiments with systems of globally coupled photochemical oscillators were performed in which two synchronized clusters emerged. In Ref. \cite{Kiss1}, the coupling function between electrochemical oscillators $n$ and $m$ was directly measured and found to be qualitatively equivalent to either $H_{nm}(\theta)=(K/N)\sin(\theta)$ at a lower voltage, which is equivalent to the classical Kuramoto model, or $H_{nm}(\theta)=(K/N)\sin(2\theta)$ at higher voltage, which is equivalent to Eq.~(\ref{eqH}) with $q=2$. In some Kuramoto-type models of neuronal networks, a coupling function of the form in Eq.~(\ref{eqH}) and the associated cluster synchrony arises as a result of learning and network adaptation. It has been proposed that the coupling between oscillators in such networks evolves according to a Hebbian learning rule \cite{Seliger1}. If this learning is fast, Eq.~(\ref{eqH}) is recovered with $q=2$ \cite{Niyogi1}. We note that the applications mentioned above all use $q=2$, but larger $q$ values are also relevant. For instance, cases of three or more clusters have appeared in the study of synthetic gene networks \cite{Zhang1}.

Cluster synchrony has been studied in many contexts, for example in networks of phase oscillators with \cite{Hansel1} and without noise \cite{Okuda1,Golomb1}, networks of integrate-and-fire oscillators \cite{Mauroy1}, and more general cases \cite{Zanette1}. 
Reference \cite{Seliger1} studied Eqs.~(\ref{eqModelGeneral}) and (\ref{eqH}) in steady-state using a self-consistent approach to characterize the phase distribution and stability of the clusters. Reference~\cite{Hansel1} studied the dynamics of clusters in ensembles of oscillators when the coupling function $H$ has two Fourier modes under the effect of small noise. Despite these and other studies \cite{Okuda1}, a complete analytical treatment of  Eqs.~(\ref{eqModelGeneral}) and (\ref{eqH}) is lacking. For example, Ref.~\cite{Seliger1} studies the steady state solution using a self-consistent approach, but does not analyze the dynamics, while Refs.~\cite{Hansel1,Okuda1} assume identical oscillators. In this paper we will use the Ott-Antonsen ansatz to obtain a low-dimensional description of cluster dynamics and a full solution to Eqs.~(\ref{eqModelGeneral}) and (\ref{eqH}). Thus, our solution of Eqs.~(\ref{eqModelGeneral}) and (\ref{eqH}) is analogous to the recent solution \cite{OA1,OA2} of the Kuramoto model in that, even though partial solutions existed previously, our solution fully characterizes the dynamics (with the same caveats as in Refs.~\cite{OA1,OA2}).

Two interesting phenomena that are particular to systems displaying cluster synchrony are asymmetric clustering \cite{Banaji1} and switching \cite{Hansel1,Taylor1}. Asymmetric clustering is characterized by a non-uniform distribution of oscillators in different clusters and switching refers to oscillators moving between clusters. We find that asymmetric clustering emerges from non-uniform initial conditions, to which systems with a coupling function of the form of Eq.~(\ref{eqH}) with $q\ge2$ are very sensitive. Switching can be achieved by introducing an external forcing term, $F\sin(\Psi-\omega_0t-\theta_n)$ (where $\omega_0$ is the average oscillator frequency), on the right hand side of Eq.~(\ref{eqModelGeneral}) with $F\ne0$ nonzero for a finite amount of time. This results in a fraction of oscillators switching to a cluster around $\theta=\Psi$. If different values of $\Psi$ are chosen for different oscillators (i.e., $\Psi\mapsto\Psi_n$), then if $F$ is large enough with respect to $|\omega_n|$ and $K$, the phase of oscillator $n$ will converge to a value near $\Psi_n$. 

This paper is organized as follows. In Section II we solve for the dynamics of the system with Eq.~(\ref{eqH}) and $q=2$. In Section III we discuss the effect of external forcing on asymmetric clustering and switching, the presence of hysteresis when the coupling strength is changed, as well as how asymmetric clustering can be used to store information. In Section IV we discuss how results generalize to the case $q>2$. Finally, in Section V we conclude this paper by discussing our results. 

\section{Low-dimensional description of the two-cluster state}

In this section, we will study in detail Eqs. (\ref{eqModelGeneral}) and (\ref{eqH}) with $q=2$, which leads to the system
\begin{equation}\label{eqq2}
\dot{\theta}_n = \omega_n+\frac{K}{N}\sum_{m=1}^N\sin[2(\theta_m-\theta_n)],
\end{equation}
where the intrinsic frequencies $\omega_n$ are drawn randomly from a distribution $g(\omega)$. Also, we define the set of generalized order parameters
\begin{equation}\label{eqGenOrderPar}
r_k=|r_k| e^{i\psi_k}=\frac{1}{N}\sum_{m=1}^Ne^{ik\theta_m},
\end{equation}
for $k\in\mathbb{N}$. These generalized order parameters were introduced in previous work \cite{Daido3} where coupling functions with higher harmonics were studied. We will see that when more than one cluster emerges, more than one $r_k$ is needed to describe the dynamics of the system. In this case of $q=2$, two clusters emerge. The order parameter $|r_2|$ measures the degree of cluster synchrony in the system while $|r_1|$ measures the degree of asymmetry in clustering (see Fig.~\ref{cartoon}). Eq.~(\ref{eqq2}) can be rewritten in terms of $r_2$ as
\begin{equation}\label{eqq2MF}
\dot{\theta}_n = \omega_n + \frac{K}{2i}\left(r_2e^{-2i\theta_n}-r_2^*e^{2i\theta_n}\right),
\end{equation}
where $^*$ denotes complex conjugate.

Following Ref.~\cite{OA1}, we let $N\to\infty$ and move to a continuum description. Accordingly, we introduce the density function $f(\theta,\omega,t)$, which describes the density of oscillators with phase $\theta$ and natural frequency $\omega$ at time $t$. Since oscillators are conserved $f$ must satisfy the continuity equation $\partial_t f+\partial_\theta(f\dot{\theta})=0$, giving
\begin{align}\label{eqPDE}
\partial_t f &+ \partial_\theta\left[f\left(\omega+\frac{K}{2i}\left(r_2e^{-2i\theta}-r_2^*e^{2i\theta}\right)\right)\right]=0.
\end{align}
To analyze Eq.~(\ref{eqPDE}), we find it convenient to define the symmetric and antisymmetric parts of $f$, $f_s$ and $f_a$, as
\begin{align}
f_{s/a}(\theta,\omega,t) &= [f(\theta,\omega,t)\pm f(\theta+\pi,\omega,t)]/2,
\end{align}
where $f_s$ and $f_a$ are symmetric and antisymmetric with respect to translation by $\pi$, respectively, in the sense that $f_s(\theta+\pi,\omega,t)=f_s(\theta,\omega,t)$ and $f_a(\theta+\pi,\omega,t)=-f_a(\theta,\omega,t)$. We note that $f$ is a solution of Eq.~(\ref{eqPDE}) if $f = f_s+f_a$ and $f_s$ and $f_a$ are both solutions of Eq.~(\ref{eqPDE}). Thus, we can study separately the symmetric and antisymmetric dynamics of solutions $f$.

\subsection{Symmetric Dynamics}

We now use a variation of the OA ansatz to find a low-dimensional analytical solution for the dynamics of the symmetric part of $f$, which evolves independently from the antisymmetric part. For the Kuramoto model, after expanding the distribution $f$ in Fourier Series,
\begin{equation}\label{eqFourier}
f(\theta,\omega,t)=\frac{g(\omega)}{2\pi}\left(1+\sum_{n=1}^\infty \widehat{f}_n(\omega,t)e^{in\theta} + c.c.\right),
\end{equation}
where $c.c.$ denotes complex conjugate terms, Ref.~\cite{OA1} introduces the ansatz $\widehat{f}_n(\omega,t)=a^n(\omega,t)$ which yields a solution for systems with sinusoidal coupling provided $a$ evolves according to a simple ordinary differential equation (ODE). Solutions of this kind turn out to form a low-dimensional, globally-attracting, invariant manifold to which solutions converge quickly given that the spread in $g(\omega)$ is non-zero \cite{OA2}. This manifold is the set of Poisson kernels,
\begin{equation}
f(\theta,\omega,t)=\frac{g(\omega)}{2\pi}\frac{1-|a|^2}{1+|a|^2-2|a|\cos(\text{arg}(a)-\theta)}.
\end{equation}

In terms of the Fourier series (\ref{eqFourier}), the symmetric part of $f$ is given by
\begin{equation}\label{eqAnsatzEven}
f_s(\theta,\omega,t)=\frac{g(\omega)}{2\pi}\left(1+\sum_{n=1}^\infty \widehat{f}_{2n}(\omega,t)e^{2in\theta} + c.c.\right).
\end{equation}
For the new system defined by Eq.~(\ref{eqq2}), we use the following variation of the OA ansatz on the symmetric part of $f$: $\widehat{f}_{2n}(\omega,t)=a^n(\omega,t)$. When Eq.~(\ref{eqAnsatzEven}) with this ansatz is inserted into Eq.~(\ref{eqPDE}) and projected onto the subspace spanned by $e^{in\theta}$, all equations reduce to the following ODE for $a$:
\begin{equation}\label{eqa}
\dot{a}+2i\omega a+K\left(r_2 a^2 - r_2^*\right)=0.
\end{equation}

In the continuum limit, we have
\begin{align}\label{eqR2Lim}
r_2(t) &= \int_{-\infty}^\infty \int_0^{2\pi} e^{2i\theta}f(\theta,\omega,t)d\theta d\omega\nonumber\\
&= \int_{-\infty}^\infty g(\omega)a^*(\omega,t)d\omega.
\end{align}

We now assume that $g(\omega)$ is Lorentzian with mean $\omega_0$ and spread $\Delta$, i.e. $g(\omega)=\frac{\Delta}{\pi(\Delta^2+(\omega-\omega_0)^2)}$. Furthermore, by entering the rotating frame $\theta\mapsto\theta+\omega_0t$ we can assume without loss of generality that $\omega_0=0$. With this choice of $g(\omega)$ we can integrate Eq.~(\ref{eqR2Lim}) exactly by closing the contour with the semicircle of infinite radius in the lower-half complex plane and evaluating $a$ at the enclosed pole (see Refs.~\cite{OA1,OA2} for the validity of this procedure):
\begin{equation}
r_2(t)=a^*(-i\Delta,t)\equiv a^*(t),
\end{equation}\label{eqR2}
where we've defined $a(t)\equiv a(-i\Delta,t)$ to simplify notation. By evaluating Eq.~(\ref{eqa}) at $\omega=-i\Delta$, close the dynamics for $r_2$:
\begin{align}\label{eqR2}
\dot{r_2} = -2\Delta r_2 + K(r_2-r_2^*r_2^2).
\end{align}
In polar coordinates, $r_2=|r_2|e^{i\psi_2}$, we find
\begin{align}
\dot{|r_2|} &= -2\Delta|r_2| + K|r_2|(1-|r_2|^2), \label{eqrho2}\\
\dot{\psi}_2 &= 0.
\end{align}
Thus, the unsynchronized state (i.e. $|r_2|=0$) is stable for $K<2\Delta$, at which point it loses stability and the stable synchronized branch $|r_2|=\sqrt{1-2\Delta/K}$ emerges.

We now show that solutions of the form given in Eq.~(\ref{eqAnsatzEven}) with $f_n(\omega,t)=a^n(\omega,t)$, where $a$ obeys Eq.~(\ref{eqa}), are globally attracting. An alternative way of solving for the dynamics of $r_2$ is to make the change of variable $\phi=2\theta$, which yields a new continuity equation:
\begin{align}
\partial_t f_s + \partial_\phi \left[2f_s\left(\omega+\frac{K}{2i}(r_2e^{-i\phi}-r_2^*e^{i\phi})\right)\right]=0,
\end{align}
which is of the same form of the equation studied in Ref.~\cite{OA2}. There it is shown that the dynamics of $r_2$ given by Eq.~(\ref{eqR2}) are globally attracting provided that the spread of $g(\omega)$ is non-zero. Thus, the globally attracting invariant manifold for $f_s$ is the set of double Poisson kernels centered at $\psi_2$, $f_s(\theta,\omega,t) = P(2\theta-\psi_2,|r_2(t)|,\omega)$, where 
\begin{equation}\label{eqManifold}
P(\theta,\rho,\omega)=\frac{g(\omega)}{2\pi}\frac{1-\rho^2}{1+\rho^2-2\rho\cos(\theta)}.
\end{equation}
Since the system is invariant to rotations $\theta\mapsto\theta+\varphi$, hereafter we assume without loss of generality that $\psi_2=0$.

\subsection{Steady-State Solution}

We first find the steady-state solutions of the system described by Eq.~(\ref{eqq2}). Recall that the symmetric and antisymmetric parts of $f$ satisfy the partial differential equation (PDE)
\begin{equation}
\partial_t f_{s/a} + \partial_\theta\left[f_{s/a}\left(\omega-K|r_2|\sin(2\theta)\right)\right].
\end{equation}
To find the steady-state solution $f_{s/a}^{ss}$, we set $\partial_t f_{s/a}^{ss}=0$. 

For $|\omega|\le K|r_2|$ we find that $f_{s/a}^{ss}/g(\omega) = c_{1,s/a}\delta(\theta-\phi(\omega))+c_{2,s/a}\delta(\theta-\phi(\omega)-\pi)$, where $\phi(\omega)$ and $\phi(\omega)+\pi$ are the stable fixed points of Eq.~(\ref{eqq2MF}) defined by $\phi(\omega)=\frac{1}{2}\arcsin[\omega/(K|r_2|)]$. (Recall that we assume $\psi_2=0$.) Imposing symmetric and antisymmetric conditions, we have that $c_{1,s}=c_{2,s}=1/2$ and $c_{1,a}=-c_{2,a}=c$ with $|c|\le1/2$.

For $|\omega|>K|r_2|$, we find that $f_s^{ss}/g(\omega) = C(\omega)/|\omega-K|r_2|\sin(2\theta)|$, where $C(\omega)=2\sqrt{\omega^2-K^2|r_2|^2}/\pi$ and $f_a^{ss} = 0$. Thus, the complete steady-state distribution of oscillators is

\begin{widetext}
\begin{align} \label{eqfss}
f^{ss}(\theta,\omega) = \left\{\begin{array}{ll} g(\omega)[(1/2 + c)\delta(\theta-\phi(\omega)) + (1/2-c)\delta(\theta-\phi(\omega)-\pi)] & \hskip4ex \text{ if } |\omega|\le K|r_2|, \\ 2 g(\omega)\sqrt{\omega^2-K^2|r_2|^2}/|\pi[\omega-K|r_2|\sin(2\theta)]| & \hskip4ex \text{ if } |\omega|>K|r_2|, \end{array}\right.
\end{align}
\end{widetext}
with $|r_2|=\sqrt{1-2\Delta/K}$. The interpretation of the different terms in Eq.~(\ref{eqfss}) is the following. For $|\omega|\le K|r_2|$ $f^{ss}$ is comprised of two delta functions representing the two clusters of phase-locked oscillators at $\theta=\phi(\omega)$ and $\phi(\omega)+\pi$. For $|\omega|>K|r_2|$ oscillators drift for all time and the second line in Eq.~(\ref{eqfss}) is their steady-state distribution.

While the symmetric part of the distribution is only dependent on the value of $K$, the antisymmetric part of the distribution depends on the free parameter $c$, which must be determined from initial conditions. Thus, different solutions with different degrees of cluster asymmetry coexist.

\begin{figure}[b]
\centering
\addtolength{\belowcaptionskip}{-6mm}
\subfigure{
\epsfig{file =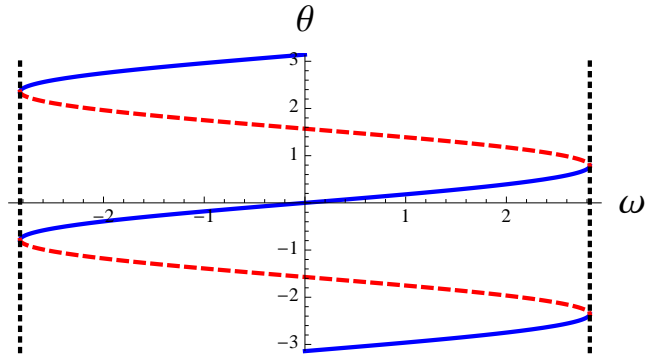, clip =,width=0.98\linewidth }
}
\caption{(Color online) Stable (solid blue) and unstable (dashed red) equilibria of $\theta$ as a function of $\omega$ for phase-locked oscillators. Boundaries between locked and drifting regions ($\omega=\pm K|r_2|$) are plotted in black dotted lines.} \label{basins}
\end{figure}

\begin{figure}[b]
\centering
\addtolength{\belowcaptionskip}{-6mm}
\subfigure{
\epsfig{file =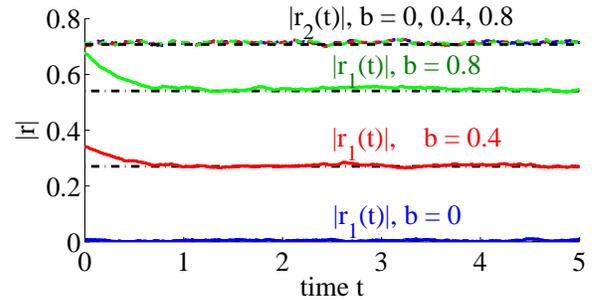, clip =,width=0.98\linewidth }
}
\caption{(Color online) Order parameters $|r_1(t)|$ (solid colored curves) and $|r_2(t)|$ (dashed colored curves) for $b=0$, $0.4$, and $0.8$ from a simulation of Eq.~(\ref{eqq2}) with $N=10000$ oscillators and analytic predictions of steady-state (black dot-dashed lines). Parameters are $K=4$, $\Delta=1$.} \label{r1a}
\end{figure}

Now we compute the degree of cluster synchrony and asymmetry in the system at steady-state in terms of initial conditions. The degree of cluster synchrony is exactly $|r_2|=\sqrt{1-2\Delta/K}$, but the degree of asymmetry, measured by $|r_1|$, depends on the free parameter $c$ which must be determined from initial conditions. To calculate $r_1$, we note that only the locked portion ($|\omega|\le K|r_2|$) of $f$ contributes to $r_1$, so 
\begin{align}
r_1&=\int_{-K|r_2|}^{K|r_2|}\int_0^{2\pi}f^{ss}(\theta,\omega)e^{i\theta}d\theta d\omega \\
&= 2c\int_{-K|r_2|}^{K|r_2|}g(\omega)e^{i\phi(\omega)}d\omega.
\end{align}
Through a series of substitutions, this integral can be evaluated exactly:
\begin{align}
|r_1| &= \frac{2\sqrt{2}c}{\pi}\left(\frac{\arctan(A^{-})}{A^{+}}-\frac{\text{arctanh}(A^{+})}{A^{-}}\right),\label{eqrhoc}\\
\text{where } & A^{\pm} = \sqrt{\frac{K|r_2|}{\sqrt{K^2|r_2|^2+\Delta^2}\pm K|r_2|}}.
\end{align}

\begin{figure}[t]
\centering
\addtolength{\belowcaptionskip}{-6mm}
\subfigure{
\epsfig{file =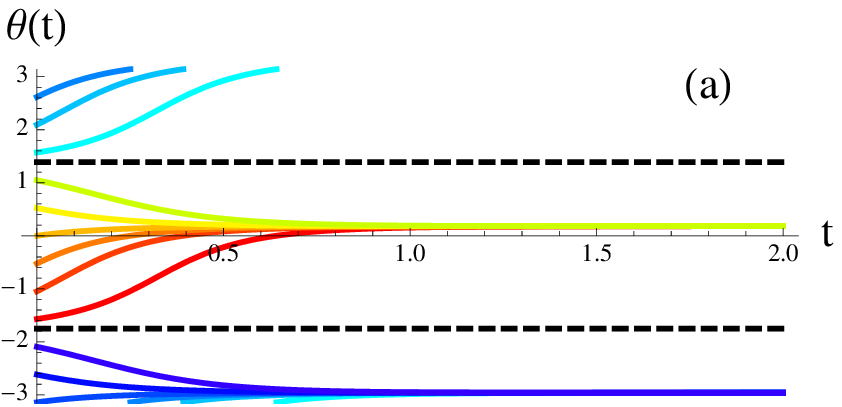, clip =,width=0.98\linewidth }
}
\subfigure{
\epsfig{file =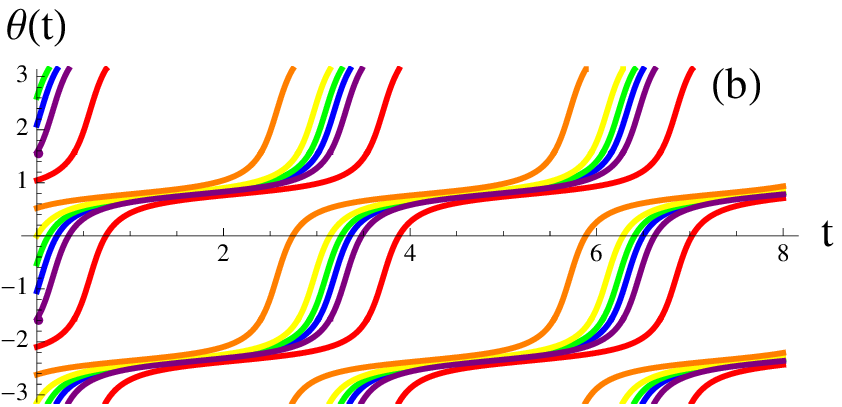, clip =,width=0.98\linewidth } 
}
\caption{(Color online) Example characteristics $\theta(t)$ from Eq.~(\ref{eqchartheta}) for $K=4$ and $\Delta=1$ of (a) locked oscillators ($\omega=1$) and (b) drifting oscillators ($\omega=3$).} \label{char}
\end{figure}

As an example illustrating the dependence of $c$ on initial conditions we assume for simplicity that the symmetric dynamics are at steady-state by time $t=t_0$ so that $|r_2|=\sqrt{1-2\Delta/K}$, but the antisymmetric part may still not be at rest. Thus, phase-locked oscillators with natural frequency $\omega_n$ settle to one of the two stable equilibria of $\dot{\theta}_n=\omega_n-K|r_2|\sin(2\theta_n)$, while the unstable equilibria serve as boundaries for the basins of attraction. In Fig.~\ref{basins} we plot the stable equilibria in blue solid lines and the unstable equilibria in red dashed lines for $K=4$ and $\Delta=1$. Boundaries between locked and drifting regions, $\omega=\pm K|r_2|$, are plotted in dotted black lines. We denote the unstable equilibria by $\Theta_{1,2}=-\frac{1}{2}\arcsin[\omega/(K|r_2|)]\mp\frac{\pi}{2}$. From Eq.~(\ref{eqfss}) we find that $c+1/2$ is just the fraction of oscillators in the locked region between $\Theta_1$ and $\Theta_2$, so $c$ in terms of the initial density $f(\theta,\omega,t_0)$ is 
\begin{equation}\label{eqc1}
c+\frac{1}{2}=\frac{\int_{-K|r_2|}^{K|r_2|}\int_{\Theta_1}^{\Theta_2}f(\theta,\omega,t_0)d\theta d\omega}{\int_{-K|r_2|}^{K|r_2|}\int_{-\pi}^{\pi}f(\theta,\omega,t_0)d\theta d\omega}.
\end{equation}

To test this result, we choose initial conditions
\begin{align}
f(\theta,\omega,t_0)=P(2\theta,\rho_2,\omega)[1+b\cos(\theta)],
\end{align}
which has symmetric and antisymmetric parts $f_s=P(2\theta,|r_2|,\omega)$ and $f_a=bP(2\theta,|r_2|,\omega)\cos(\theta)$, respectively. 

Choosing $K=4$ and $\Delta=1$ we integrate Eq.~(\ref{eqc1}) numerically to get $c\approx0.442351 b$. In Fig.~\ref{r1a} we plot results from a direct numerical simulation of Eq.~(\ref{eqq2}) compared with the analytical prediction above. We simulate $N=10000$ oscillators with $K=4$ and $\Delta=1$ and plot $|r_1(t)|$ for $b=0$, $0.4$, and $0.8$ in blue, red, and green solid lines (labeled in Fig.~\ref{r1a}), respectively, with the predicted value of $\lim_{t\to\infty}|r_1(t)|$ for each in black dot-dashed. We also plot $|r_2(t)|$ for each value and the predicted value of $\lim_{t\to\infty}|r_2(t)|=1/\sqrt{2}$ in black dashed curves. Simulations agree very well with the theory. Note that, unlike $|r_1|$, $|r_2|$ (both predicted and from simulation) does not depend on $b$.

\subsection{Transient Dynamics}

From Fig.~\ref{r1a} we see that the $|r_1|$ dynamics reach steady-state quickly. To capture the transient dynamics we can solve the PDE (\ref{eqPDE})
\begin{equation}\label{eqPDE2}
\partial_t f + [\omega-K|r_2|\sin(2\theta)]\partial_\theta f = 2K|r_2| \cos(2\theta)f
\end{equation}
coupled with the $|r_2|$ dynamics, which evolve independently, via the method of characteristics \cite{Guenther1}. The characteristic equations (along with $\dot{\omega}=0$) are
\begin{align}
\dot{\theta} &= \omega-K|r_2|\sin(2\theta), \label{eqchar1} \\
\dot{f} &= 2K|r_2| \cos(2\theta), \label{eqchar2} \\
\dot{|r_2|} &= 2\left[-\Delta|r_2| + \frac{K}{2}|r_2|(1-|r_2|^2)\right]. \label{eqchar3}
\end{align}

\begin{figure}[t]
\centering
\addtolength{\belowcaptionskip}{-6mm}
\subfigure{
\epsfig{file =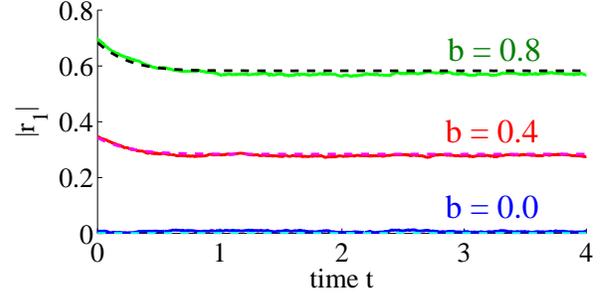, clip =,width=0.98\linewidth }
}
\caption{(Color online) Transient dynamics of $|r_1(t)|$ for initial conditions $f(\theta,\omega,0)=P(2\theta,|r_2|,\omega)(1+b\cos(\theta))$ from simulation with $N=10000$ and $b=0$, $0.4$, and $0.8$ (blue, red and green curves) and from integrating Eq.~(\ref{eqr1int2}) numerically (cyan, magenta, and black dashed curves). Parameters are $K=4$ and $\Delta=1$.} \label{trans}
\end{figure}

When $|r_2|$ is at steady-state Eqs.~(\ref{eqchar1}-\ref{eqchar3}) can be solved analytically. Analytic expressions for the characteristic curves $\theta(t,\theta_0)$ starting at the initial phase $\theta_0$ and the distribution $f(\theta,\omega,t)$, starting with initial condition $f(\theta,\omega,t_0)=g(\omega)h(\theta)$ are given in Appendix A. 

\begin{figure*}[t]
\centering
\addtolength{\belowcaptionskip}{-6mm}
\subfigure{
\epsfig{file =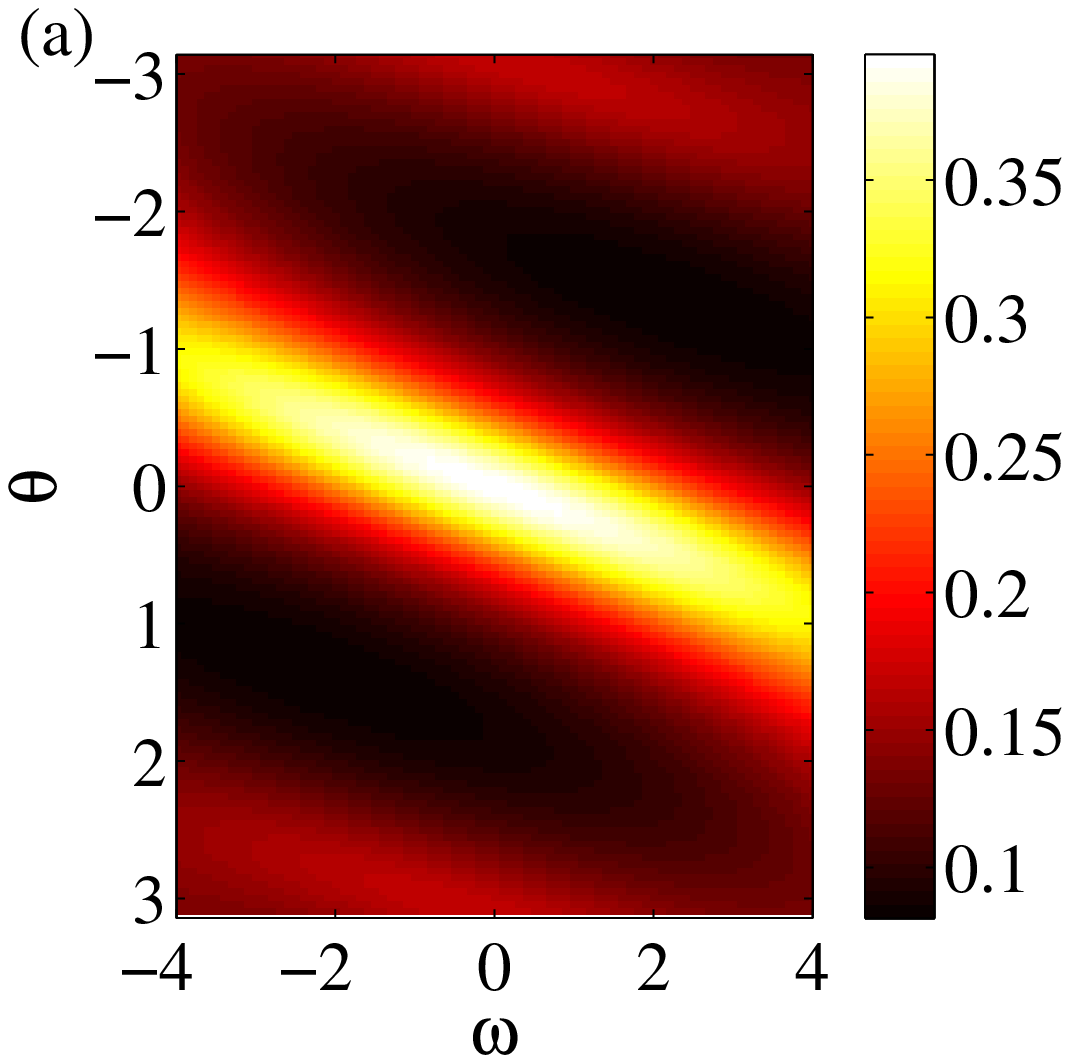, clip =,width=0.32\linewidth }
}
\subfigure{
\epsfig{file =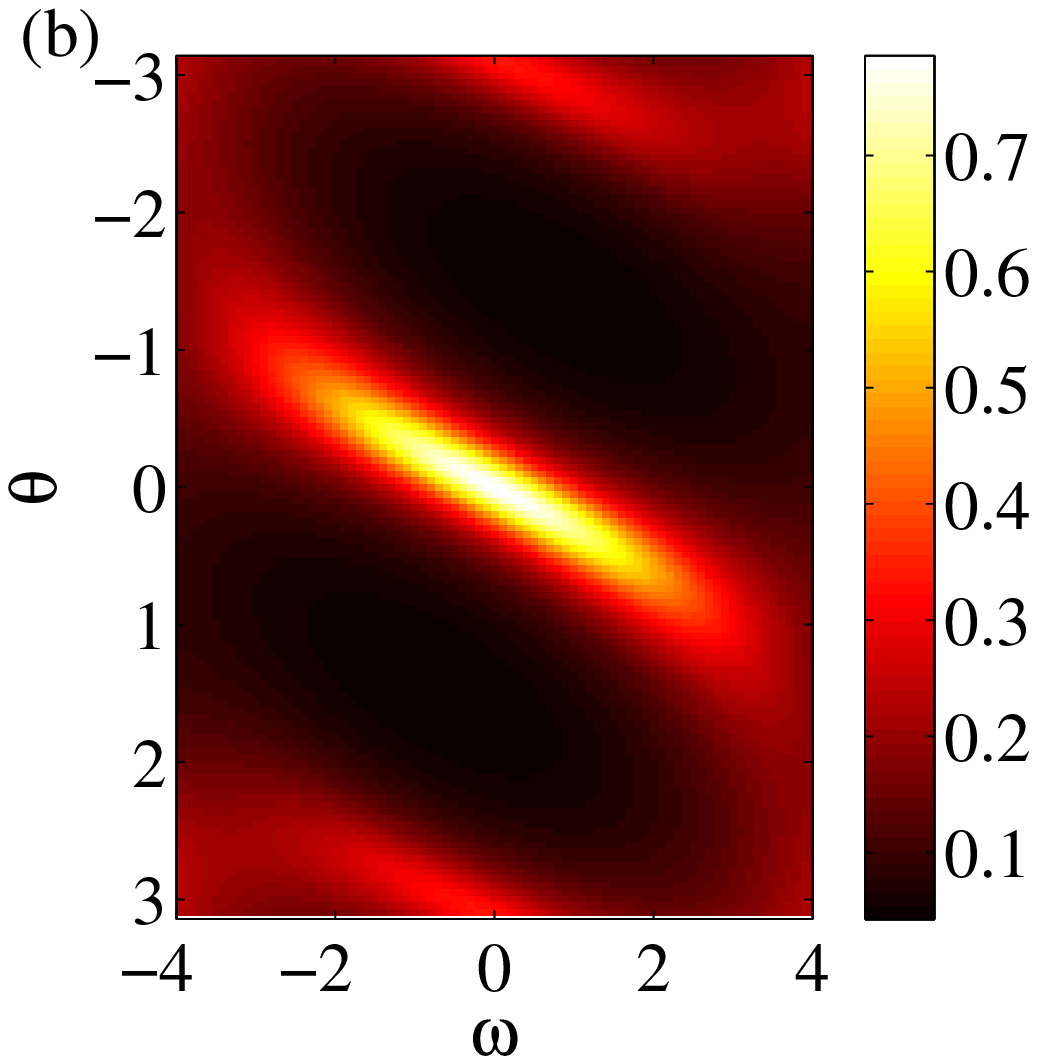, clip =,width=0.32\linewidth } 
}
\subfigure{
\epsfig{file =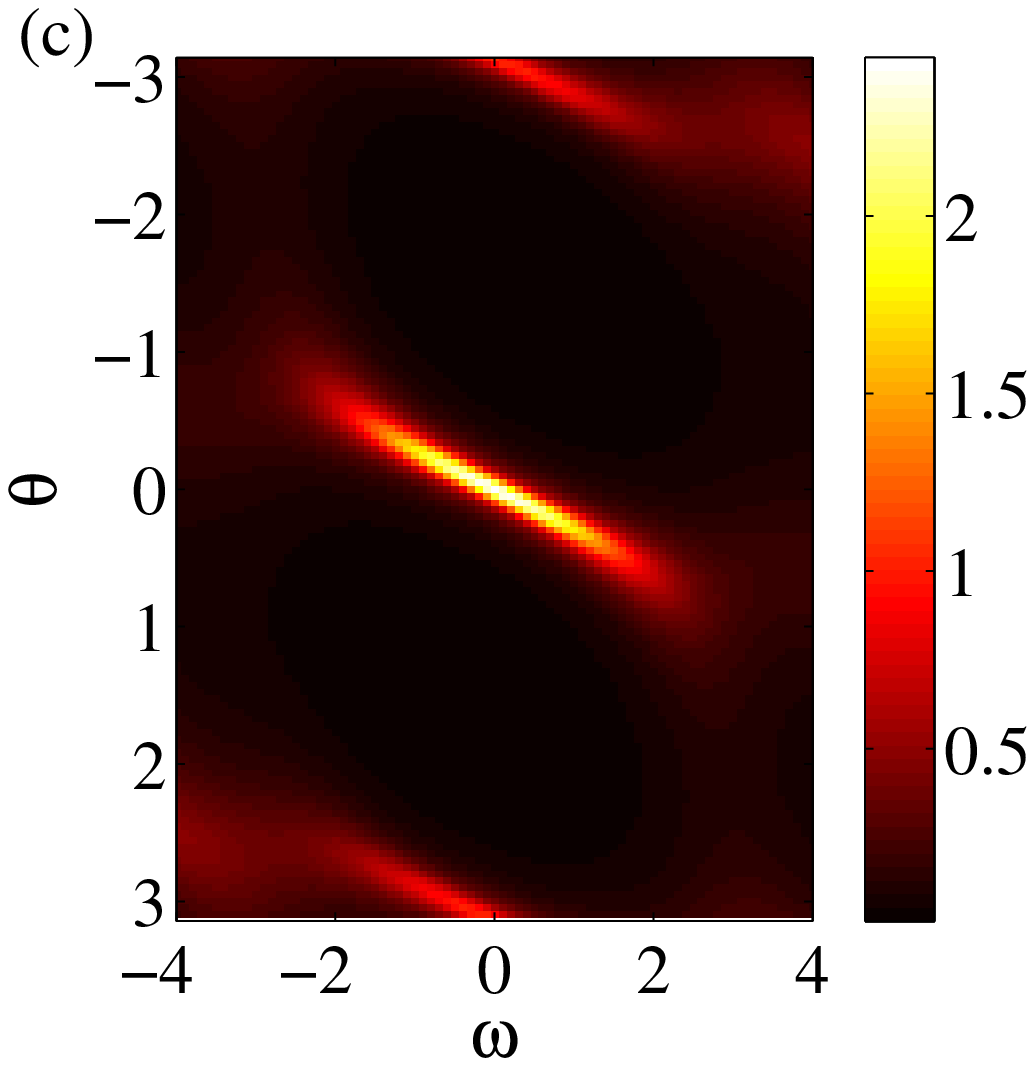, clip =,width=0.32\linewidth } 
}
\caption{(Color online) Numerically-computed distribution $f(\theta,\omega,t)/g(\omega)$ obtained from numerically solving Eq.~(\ref{eqPDE2}) at times $t=0.33$ (a), $t=0.67$ (b), and $t=1$ (c) with inital conditions $P(2\theta,0.1,\omega)(1+0.4\cos(\theta))$ and parameters $K=4$ and $\Delta=1$.} \label{fPDE}
\end{figure*}

\begin{figure}[b]
\centering
\epsfig{file =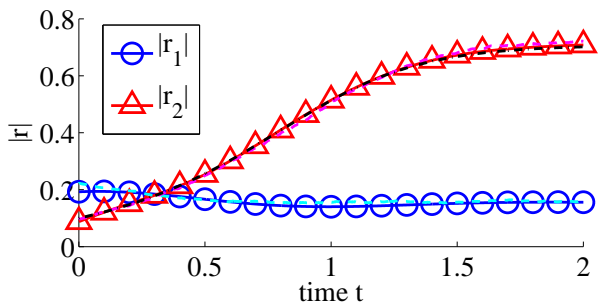, clip =,width=1.0\linewidth }
\caption{(Color online) Comparison of $|r_1|$ and $|r_2|$ from the numerically solved PDE (\ref{eqPDE2}) (blue circles and red triangles) and from direct simulation of Eq.~(\ref{eqq2}) with $N=10000$ oscillators (cyan and magenta dashed lines) with the same initial conditions and parameters used in Fig.~\ref{fPDE}.}\label{rPDESim}
\end{figure}

Using $K=4$ and $\Delta=1$, we plot some example characteristics using the analytic solution for $\omega=1$ and $\omega=3$ in Figs.~\ref{char}(a) and \ref{char}(b), respectively. For these parameter values $\omega=1$ is in the locked population and $\omega=3$ is drifting. For $\omega=1$, characteristics (solid colored lines) quickly converge to one of the two stable fixed points, with basins of attraction separated by unstable fixed points (black dotted lines). Thus, $f$ evaluated at $\omega=1$ converges very quickly to two point masses. However, for $\omega=3$, the characteristics continue drifting with a finite velocity for all time.

In principle, we could calculate $r_1(t)$ through the integral
\begin{equation}\label{eqr1int1}
r_1(t) = \int_{-\infty}^\infty\int_{-\pi}^\pi f(\theta,\omega,t)e^{i\theta}d\theta d\omega,
\end{equation}
where $f(\theta,\omega,t)$ is given by Eq.~(\ref{eqftrans}) in Appendix A. However, given the quick convergence of $f$ to delta functions in the locked regime, Eq.~(\ref{eqr1int1}) is difficult to integrate numerically, so we rather calculate $r_1(t)$ via the integral
\begin{equation}\label{eqr1int2}
r_1(t)=\int_{-\infty}^\infty\int_{-\pi}^\pi f(\theta(t,\theta_0),\omega,t)e^{i\theta(t,\theta_0)}\frac{\partial\theta}{\partial\theta_0}d\theta_0 d\omega.
\end{equation}

In Fig.~\ref{trans} we compare the results of integrating Eq.~(\ref{eqr1int2}) numerically with the simulations of Eq.~(\ref{eqq2}) using $N=10000$ oscillators, $K= 4$, $\Delta=1$, and $f(\theta,\omega,0)=P(2\theta,|r_2|,\omega)(1+b\cos(\theta))$. For $b = 0$, $0.4$, and $0.8$, $|r_1|$ obtained from simulations are plotted as solid lines, and results from integrating Eq.~(\ref{eqr1int2}) numerically are plotted as dashed lines. The results from Eq.~(\ref{eqr1int2}) capture the transient dynamics very well.

The example above leading to Fig. \ref{trans} was for a case with $|r_2|$ initially at steady state. If $|r_2|$ is not initially at steady-state, the solution to Eq.~(\ref{eqrho2}) with initial condition $|r_2(0)|=\rho_0$ is exactly \cite{OA1}
\begin{align}\label{eqp2}
|r_2(t)| = \overline{P}_2/\sqrt{1+\left[\left(\frac{\overline{P}_2}{\rho_0}\right)^2-1\right]e^{2(2\Delta-K)t}},
\end{align}
where $\overline{P}_2=\sqrt{1-2\Delta/K}$.

In Fig.~\ref{fPDE} we plot the evolution of $f(\theta,\omega,t)$ obtained from numerically solving Eq.~(\ref{eqPDE2}) when the symmetric dynamics are not at steady-state. Starting with initial conditions $f(\theta,\omega,0)=P(2\theta,0.1,\omega)(1+0.4\cos(\theta))$ and parameters $K=4$, $\Delta = 1$ we plot the distribution$f(\theta,\omega,t)/g(\omega)$ at $t=0.33$ (a), $t=0.67$ (b), and $t=1$ (c). We see that the distribution quickly localizes, in agreement with the asymptotic form in Eq.~(\ref{eqfss}). In Fig.~\ref{rPDESim} we compare $|r_1(t)|$ and $|r_2(t)|$ calculated from the numerical solution of Eq.~(\ref{eqPDE2}) (blue circles and red triangles) with the same variables calculated from a direct simulation of Eq.~(\ref{eqq2}) with $N=10000$ oscillators (cyan and magenta dashed lines). The analytic solution $|r_2(t)|$ in Eq.~(\ref{eqp2}) is plotted as a black dot-dashed line.

\section{External Driving and Hysteresis in the Two-Cluster State}

As we have seen, Eq.~(\ref{eqq2}) admits a family of steady-state solutions characterized by a free parameter $c$. In this Section, we demonstrate that, by appropriately forcing Eq.~(\ref{eqq2}) and modulating the coupling strength, the system can be driven to any of these solutions, thus allowing us to encode any desired value of $c$ in the state of the system. Assuming $\omega_0=0$, we consider the forced system
\begin{align}\label{eqModelForce}
\dot{\theta}_n &= \omega_n +\frac{K}{N}\sum_{m=1}^N\sin[2(\theta_m-\theta_n)] + F(t)\sin(\Phi_n-\theta_n)\\
&\text{where } F(t) = \left\{\begin{array}{rl}F_0 & \text{ if } t\in I \\ 0 & \text{ otherwise,}\end{array}\right.
\end{align}
for some forcing magnitude $F_0$ and time interval $I=[t_1,t_2]$.

For $F_0$ sufficiently large in comparison with $|\omega_n|$ and $K$ and duration $d=t_2-t_1$ not too small, $\theta_n$ will approach $\approx \arcsin(\omega_n/F_0) + \Phi_n\approx \Phi_n$ for $F_0\gg|\omega_n|+K$. Thus, if $\Phi_n=0$ for all $n$ with $d$ and $F_0$ large enough (i.e. $F_0\gg K\sqrt{1-2\Delta/K}$), all locked oscillators are entrained to the $\theta=0$ cluster and remain there after $t=t_2$, thus creating a completely asymmetric cluster state. On the other hand, if $\Phi_n$ are drawn from the distribution $h(\Phi) = d\delta(\Phi) + (1-d)\delta(\Phi-\pi)$, then the ratio of the number of oscillators ending up in the cluster centered at $\psi=0$ to those in the cluster centered at $\psi=\pi$ is $d/(1-d)$, which forces $c$ in Eq.~(\ref{eqrhoc}) to $d-1/2$. Thus, by choosing appropriately the external forcing, we can set any degree of asymmetry we wish.

\begin{figure}[t]
\centering
\addtolength{\belowcaptionskip}{-6mm}
\subfigure{
\epsfig{file =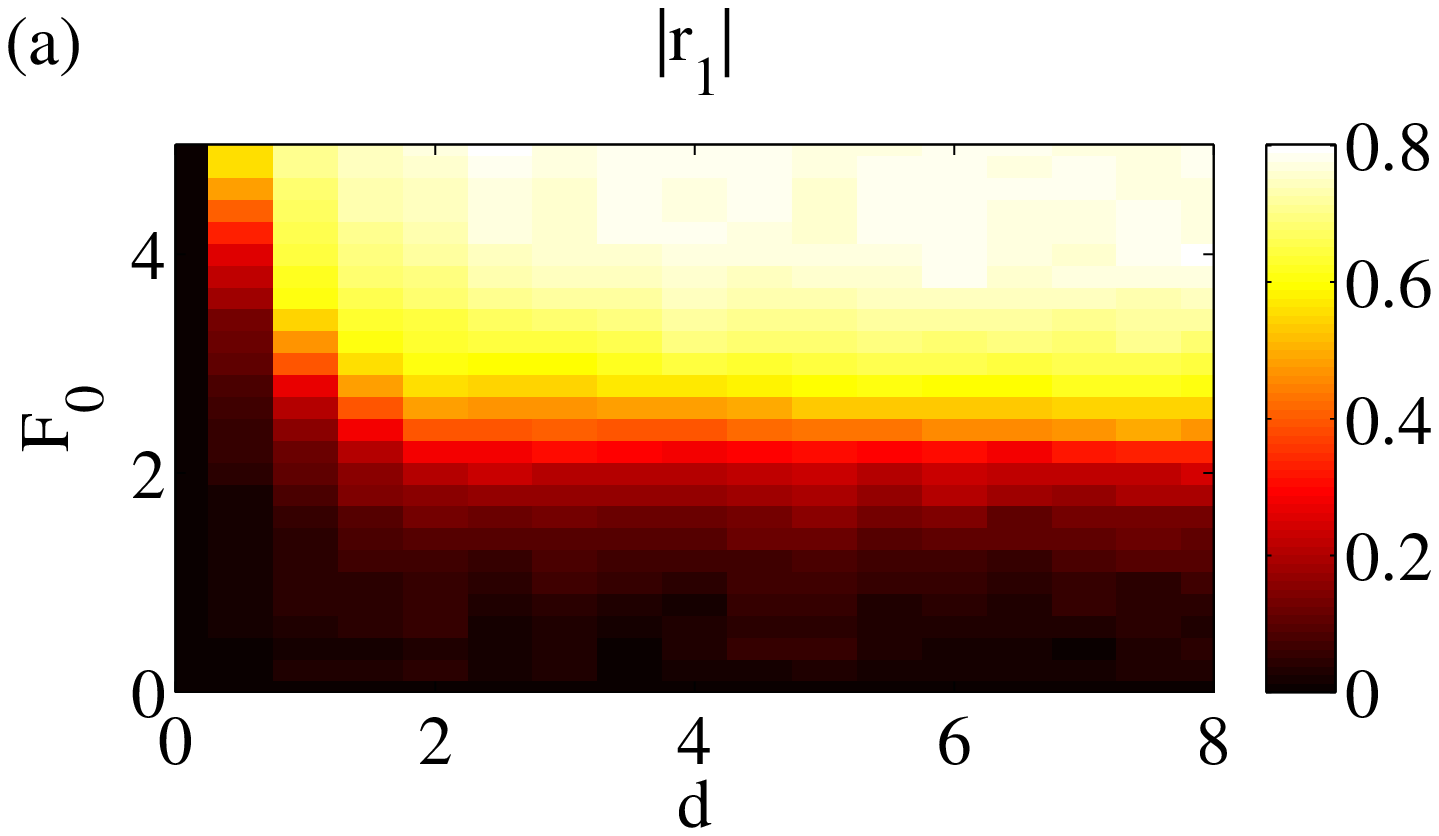, clip =,width=0.98\linewidth }
}
\subfigure{
\epsfig{file =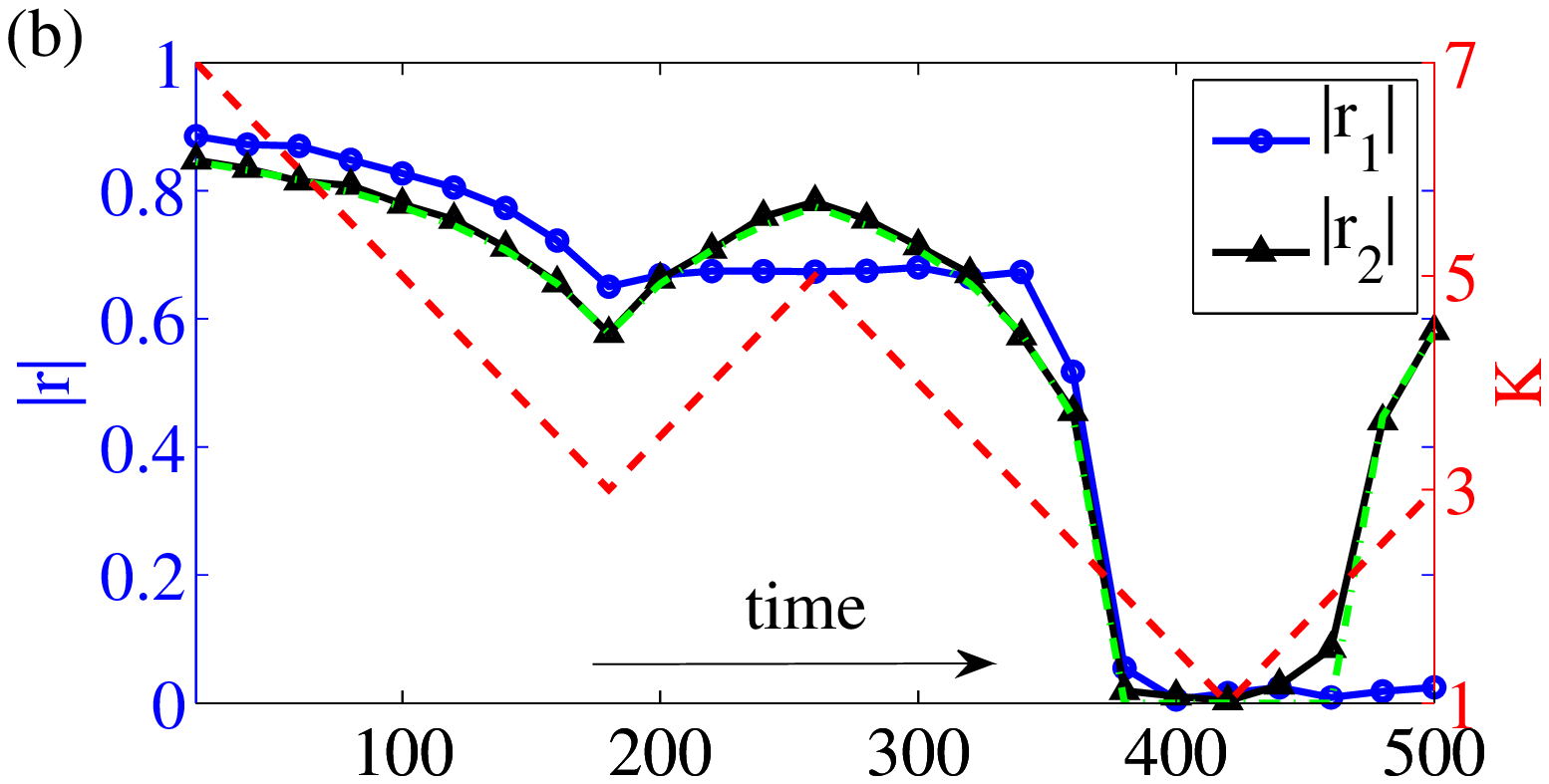, clip =,width=0.98\linewidth } 
}
\caption{(Color online) (a) Steady-state $|r_1|$ after forcing a symmetric distribution with forcing magnitude $F_0$ for a duration $d$ with $K=4$ and $\Delta=1$. (b) Hysteretic behavior of $|r_1|$ (blue circles) vs the non-hysteretic behavior of $|r_2|$ (black triangles) when $K$ is changed in time (red dashed line).} \label{forcehyst}
\end{figure}

To explore the effect of different $F_0$ and $d$ values, we simulate Eq.~(\ref{eqModelForce}) with $N = 2000$ oscillators with random initial conditions and parameters $K=4$ and $\Delta=1$ until steady-state (and attaining two clusters of approximately equal size, $|r_1|\approx0$), then force the system with a strength of $F_0$ for a duration $d$ and all $\Phi_n=0$, then allow the system to reach steady-state and plot the resulting $|r_1|$ value in Fig. \ref{forcehyst}(a). For very small $F_0$ or $d$, $|r_1|$ remains small, but as soon as both are large enough the resulting $|r_1|$ increases quickly. 

By forcing the system in this manner we achieve switching, i.e. oscillator $n$ switches to the cluster centered at phase $\Phi_n$ if $\omega_n$ is not too large. We note here that this kind of forced switching is qualitatively different than that in Ref. \cite{Taylor1}. In our original system given by Eq.~(\ref{eqq2}), switching does not occur spontaneously. Thus, external forcing is necessary to observe the phenomenon. However, in Ref. \cite{Taylor1} switching occurs spontaneously due to a heteroclinic orbit between different cluster states. 

Next, we consider the effects of slowly (compared with $\Delta^{-1}$) changing the coupling strength $K$ after a steady state with some asymmetry is reached. If steady state is reached at $t=t_0$ with a coupling strength $K=K_0$, then consider changing $K$ to $K_1$. We find hysteretic behavior in $|r_1|$ but not $|r_2|$. Regardless of whether $K_1<K_0$ or vice-versa, $|r_2|$ converges quickly to the predicted value $|r_2|=\sqrt{1-2\Delta/K_1}$, but the dynamics of $|r_1|$ are more interesting: if $K_1<K_0$ then $|r_1|$ decreases significantly, but if $K_1>K_0$, then $|r_1|$ remains approximately constant. In this situation, at time $t_0$, the distribution of oscillators is given by Eq.~(\ref{eqfss}). If $K_1<K_0$ the locked population loses all oscillators with $\sqrt{K_1^2-2\Delta K_1}<|\omega|<\sqrt{K_0^2-2\Delta K_0}$ and $|r_1|$ changes accordingly (maintaining the same $c$ value, since these oscillators are lost in equal proportions from both clusters). On the other hand, if $K_1>K_0$ the locked population will gain oscillators with $\sqrt{K_0^2-2\Delta K_0}<|\omega|<\sqrt{K_1^2-2\Delta K_1}$. However, at $t=t_0$ the distribution for these drifting oscillators is perfectly symmetric, so both clusters pick up an equal number of oscillators and the symmetric density $f_s$ changes, while the antisymmetric density $f_a$ remains the same. Thus, the only change in $|r_1|$ comes from the slight tightening of the phases $\phi(\omega)=\frac{1}{2}\arcsin[\omega/(K|r_2|)]$ and $\phi(\omega)+\pi$ about the clusters at $\theta=0$ and $\pi$.

Extending this analysis to the case where $K$ is both increased and decreased, $|r_1|$ will never increase significantly, and only decrease significantly when $K$ is decreased below a previous minimum. In figure \ref{forcehyst}(b) we plot $|r_1|$ and $|r_2|$ (blue circles and black triangles, respectively) as we change $K$ (red dashed line). While $|r_2|$ follows the predicted behavior (green dot-dashed line) without any hysteresis, it is clear that $|r_1|$ behaves as described above.

We now suggest, as others have \cite{Ashwin1}, that systems such as that given by Eq.~(\ref{eqq2}) provide ways for encoding and storing data. These systems have the unique property that the symmetric dynamics have a unique, (globally) stable fixed point, while there is a high degree of multi-stability in the antisymmetric dynamics. Furthermore, we have demonstrated above that through forced switching and modulation of the coupling strength, the asymmetry (i.e., $|r_1|$) can be controlled. Thus, we suggest that a continuous valued variable could be stored and retrieved by representing it by $|r_1|$. Furthermore, in the general $q$ case, which we study next, we will see that in addition to one globally-attracting symmetric part, there are $q-1$ additional modes that display multi-stability. Thus, through similar techniques the $q-1$ quantities $|r_1(t)|,\dots,|r_{q-1}(t)|$ can be controlled and used to store and retrieve $q-1$ different continuous valued variables. 

\section{General $q\ge2$}

We now discuss how the dynamics of the two state case generalize to higher-order coupling functions. Thus, we study the system
\begin{align}
\dot{\theta}_n & = \omega_n+\frac{K}{N}\sum_{m=1}^N\sin[q(\theta_m-\theta_n)] \nonumber \\
&= \omega_n + \frac{K}{2i}\left(r_qe^{-qi\theta_n }- r_q^*e^{qi\theta_n}\right),\label{eqq}
\end{align}
for integer $q\ge 2$ and $\omega_n$ randomly drawn from the distribution $g(\omega)$. We find in this situation that $q$ clusters form.

Again, we introduce a continuum description and represent the distribution of oscillators with the density function $f(\theta,\omega,t)$, which satisfies the continuity equation
\begin{align}\label{eqcontq}
\partial_t f + \partial_\theta\left[f\left(\omega+\frac{K}{2i}\left(r_qe^{-qi\theta}-r_q^*e^{qi\theta}\right)\right)\right]=0.
\end{align}
In analogy with Eq.~(\ref{eqAnsatzEven}) we define the modes
\begin{align}
f_j(\theta,\omega,t) &= \frac{1}{q}\sum_{k = 0}^{q-1}f(\theta+2k\pi/q,\omega,t)\exp(2\pi i j k \theta/q),
\end{align}
for $j = 0,\dots,q$. These modes satisfy the symmetry relation $f_j(\theta+2\pi/q,\omega,t)=\exp(2\pi i j\theta/q)f_j(\theta,\omega,t)$.

In analogy with the $q=2$ state, we will find that the mode $j=0$, corresponding to the symmetric part of $f$ when $q=2$, has a globally-attracting low-dimensional description that evolves independently from the other modes, leaving $q-1$ free parameters to describe the distribution.

\subsection{Dynamics of the $j=0$ Mode}

A similar variation of the OA ansatz can be used to find a low-dimensional description of the dynamics of the $j=0$ mode dynamics. The ansatz
\begin{align}\label{eqansatzq}
f_0(\theta,\omega,t)=\frac{g(\omega)}{2\pi}\left(1+\sum_{n=1}^\infty a^n(\omega,t)e^{qin\theta} + c.c.\right),
\end{align}
yields the following ODE for $a$:
\begin{equation}
\dot{a}+q\left(i\omega a + \frac{K}{2}\left(r_qa^2-r_q^*\right)\right)=0.
\end{equation}

As before, we let $g(\omega)$ be Lorentzian with zero mean and spread $\Delta$, such that $r_q(t) = a^*(-i\Delta,t)\equiv a^*(t)$, which closes the dynamics for $r_q = |r_q| e^{i\psi_q}$:
\begin{align}
\dot{|r_q|} &= q\left(-\Delta|r_q| + \frac{K}{2}|r_q|(1-|r_q|^2)\right),\\
\dot{\psi}_q &= 0.
\end{align}
Thus, the manifold for the $j=0$ mode dynamics, which can be shown to be globally attracting \cite{OA2,Ott1}, is the set of $q$-tuple Poisson kernels $P(q\theta-\psi_q,|r_q(t)|,\omega)$. Again, we assume without loss of generality that $\psi_q=0$.

\subsection{Steady-State Solution}
With $q$ potential clusters, the order-parameter $|r_q|$ measures the degree of cluster synchrony in the system, while the lower $q-1$ order parameters $|r_1|,\dots,|r_{q-1}|$ measure the degree of asynchrony. Note that the distribution is only perfectly symmetric if $r_1=\dots=r_{q-1}=0$. Thus, there are $q-1$ different measures of the asymmetry. 

Using a similar analysis as in the $q=2$ case, we find that at steady-state
\begin{widetext}
\begin{align}
f^{ss}(\theta,\omega) = \left\{\begin{array}{ll} g(\omega)\sum_{j=0}^{q-1}(1/q+c_j)\delta(\theta-\phi(\omega)-2k\pi/q) & \hskip4ex \text{ if }  |\omega|\le K|r_q|, \\ q g(\omega)\sqrt{\omega^2-K^2|r_q|^2}/|\pi[\omega-K|r_q|\sin(q\theta)]| & \hskip4ex \text{ if } |\omega|>K|r_q|, \end{array}\right.
\end{align}
\end{widetext}
with $|r_q|=\sqrt{1-2\Delta/K}$ and $\phi(\omega) = \arcsin\left(\frac{\omega}{K|r_q|}\right)/q$. Note that for $f^{ss}$ to be a distribution the coefficients $c_j$ must satisfy $c_j\ge-1/q$ and $\sum_{j=0}^{q-1}c_j=0$, leaving $q-1$ free parameters that define the distribution. Note that in the $q=2$ case there was a single parameter [i.e., $c$ in Eq.~(\ref{eqfss})] that characterized the asymmetry between the two clusters.

\begin{figure}[t]
\centering
\epsfig{file =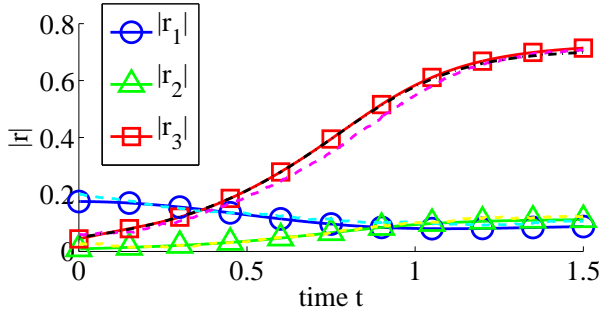, clip =,width=1.0\linewidth }
\caption{(Color online) Comparison of $|r_1|$, $|r_2|$, and $|r_3|$ from the PDE (\ref{eqfPDEq}) (blue circles, green triangles and red squares) to simulation with $N=10000$ oscillators (cyan and magenta dashed lines) with the same initial conditions and parameters from figure \ref{fPDE}.}\label{r3PDESim}
\end{figure}

The steady-state order parameters can be calculated using the same methods that led to Eq.~(\ref{eqc1}), and analogous expressions (not presented here) can be obtained.

\subsection{Transient Dynamics}
To capture the transient dynamics, we study the PDE and corresponding characteristics given by
\begin{align}
\partial_t f + [\omega -&K|r_q|\sin(q\theta)]\partial_\theta f = qK|r_q|\cos(q\theta)f,\label{eqfPDEq}\\
 \Rightarrow \qquad \dot{\theta} &= \omega-K|r_q|\sin(q\theta), \label{eqfPDEq1}\\
  \qquad \dot{f} &= qK|r_q|\cos(q\theta), \label{eqfPDEq2}\\
  \qquad \dot{|r_q|} &= q\left[-\Delta|r_q| + \frac{K}{2}|r_q|\left(1-|r_q|^2\right)\right]. \label{eqfPDEq3}
\end{align}

When $|r_q|$ is at steady-state, we can solve Eqs.~(\ref{eqfPDEq1}) and (\ref{eqfPDEq2}) exactly, yielding equations analogous to Eqs.~(\ref{eqchartheta}) and (\ref{eqftrans}) in Appendix A for the characteristics of $\theta$ and solution $f$, which we do not present here. 

When $|r_q|$ is not at steady state its evolution is given by
\begin{equation}
|r_q(t)| = \overline{P}_q/\sqrt{1+\left[\left(\frac{\overline{P}_q}{\rho_0}\right)^2-1\right]e^{q(2\Delta-K)t}},
\end{equation}
where $\overline{P}_q=\sqrt{1-2\Delta/K}$ and Eq.~(\ref{eqfPDEq}) can be solved numerically. In Fig.~\ref{r3PDESim} we compare $|r_1|$, $|r_2|$, and $|r_3|$ from the numerically-computed PDE solution (blue circles, green triangles, and red squares, respectively) to a numerical simulation of Eq.~(\ref{eqq}) with $q=3$ and $N=10000$ oscillators (cyan, yellow, and magenta dashed lines, respectively). The analytic solution for $|r_3|$ is plotted as a dot-dashed black line.

\section{Discussion}

We have found an analytic description of both steady-state and transient dynamics of a system that shows cluster synchrony given by Eqs.~(\ref{eqModelGeneral}) and (\ref{eqH}). In the large $N$ limit, $q=2$ solutions can be decomposed into symmetric and antisymmetric parts. The symmetric part, which evolves independently from the antisymmetric part and toward a steady-state independent of initial conditions, can be found using a variation of the OA ansatz \cite{OA1} and is globally attracting. The antisymmetric part, however, is shaped by the evolution of the symmertic part, is strongly dependent on initial conditions, and has a large degree of multistability. 

We have demonstrated how to manipulate the degree of asymmetry in the cluster states through the application of a short duration forcing term and modulation of the coupling strength. Starting from a symmetric state, any degree of asymmetry can be established by choosing the appropriate duration and strength of the forcing term. Furthermore, reducing the coupling strength decreases the amount of asymmetry in the cluster configuration, while increasing it does not have the opposite effect, as shown in Fig.~\ref{forcehyst}(a). Therefore, modulations of the coupling strength can be used to ``erase'' information. While we demonstrated this procedure using a system with $q=2$, similar methods could be employed for $q > 2$. In particular, $q-1$ parameters describe the cluster configuration, and the system could be driven to a configuration that encodes desired values of these parameters by the application of appropriately chosen forcing functions. Using these techniques, it is possible to encode information in the state of the system, which might find applications in the development of Kuramoto-type neural models. 

Problems that remain open include generalization such as the presence of noise and coupling functions with two or more harmonics. Thus far the work of Ott and Antonsen \cite{OA1} has not been generalized to these cases and no low dimensional analytic solution has been found. However, we hypothesize that when noise is added to Eq.~(\ref{eqq2}) spontaneous switching can occur. The case where the coupling function has more than one harmonic has also been considered \cite{Daido1}. In certain cases, e.g. $H_{nm}(\theta)=(K_1\sin(q_1\theta)+K_2\sin(q_2\theta))/N$ where $K_2\gg K_1$, the resulting system is well-approximated to the class of systems studied in this paper and results, such as clustering and asymmetry, are qualitatively similar. 

Acknowledgements: The work of J. G. R. was supported by NSF grant No. DMS-0908221. The work of E. O. was supported by ONR grant No.  N 0014-07-0734.

\begin{appendix}

\section{Characteristics}

In this appendix we present the results of solving the PDE in Eq.~(\ref{eqPDE2}) via the method of characteristics when $|r_2|$ is at steady-state (i.e. $|r_2|=\sqrt{1-2\Delta/K}$). The characteristic ODEs are Eqs.~(\ref{eqchar1}) and (\ref{eqchar2}). Given an initial phase $\theta_0$, the $\theta$ characteristics evolve as
\begin{widetext}
\begin{align} \label{eqchartheta}
\theta(t,\theta_0) &= \arctan\left[\frac{K|r_2|-\sqrt{\omega^2-K^2|r_2|^2}\tan\left(\arctan\left[\frac{K|r_2|-\omega\tan\theta_0}{\sqrt{\omega^2-K^2|r_2|^2}}\right]-t\sqrt{\omega^2-K^2|r_2|^2}\right)}{\omega}\right].
\end{align}
\end{widetext}
Several example characteristics for the locked  and drifting populations ($\omega=1$ and $\omega=3$, respectively), are plotted in Fig.~\ref{char} (a) and (b). 

For initial conditions $f(\theta,\omega,t_0)=g(\omega)h(\theta)$ the $\theta$ characteristics can be used to solve for $f(\theta,\omega,t)$, given by
\begin{align}\label{eqftrans}
f(\theta,\omega,t) &= g(\omega)\frac{h(\theta_0)}{B^D|_{\theta=\theta_0}}B^D,
\end{align}
where $B = \omega-K|r_2|\sin[2\theta(t)]$, and
\begin{widetext}
\begin{align}
D &= \left[\omega^2+K^2|r_2|^2\cos\left(E\right)-K|r_2|\sqrt{\omega^2-K^2|r_2|^2}\sin\left(E\right)\right]\left[\frac{\omega-K|r_2|\sin[2\theta(t)]}{\omega K^2|r_2|^2-\omega^3}\right],
\end{align}
\end{widetext}
where $E = 2\arctan\left[(K|r_2|-\omega\tan\theta(t))/\sqrt{\omega^2-K^2|r_2|^2}\right]$.

\end{appendix}

\bibliographystyle{plain}


\end{document}